\documentclass[]{spie}

\usepackage{graphicx}
\begin{document}

\title{Spatially resolved enhancement of fluorescence and Raman scattering by Ag nanoparticle arrays}

\author{Nic Cade, Tom Ritman-Meer, and David Richards
\skiplinehalf Department of Physics, King's College London, Strand, London WC2R 2LS, UK}

\authorinfo{E-mail: nicholas.cade@kcl.ac.uk}

\maketitle

\begin{abstract}
Highly ordered periodic arrays of silver nanoparticles have been fabricated which exhibit surface plasmon resonances in the visible spectrum. We
demonstrate the ability of these structures to alter the fluorescence properties of vicinal dye molecules by providing an additional radiative
decay channel. Using fluorescence lifetime imaging microscopy, we have created high resolution spatial maps of the molecular lifetime
components; these show an order of magnitude increase in decay rate from a localized volume around the nanoparticles, resulting in a
commensurate enhancement in the fluorescence emission intensity. Spatial maps of the Raman scattering signal from molecules on the nanoparticles
shows an enhancement of more than 5 orders of magnitude.
\end{abstract}

\keywords{Fluorescence, lifetime, plasmon, nanoparticle, Raman, SERS}

\section{INTRODUCTION}
\label{sec:intro}

Fluorescence spectroscopy and lifetime measurements are powerful tools in the study of biological samples;\cite{festy07} however, these
techniques suffer from limited sensitivity at low fluorophore concentrations typically employed. The presence of a metallic surface can
dramatically alter the emission properties of a locally situated fluorophore, resulting in enhanced fluorescence emission and greater
photostability by reducing the excited state lifetime.\cite{barnes1998,lakowicz2002,levequefort2007} Plasmon-induced fluorescence modifications
are currently being investigated for a wide variety of nanostructured systems, such as metal-island films,\cite{gheddes2007} lithographically
patterned nanoparticle arrays,\cite{gerber07} and individual gold nanospheres.\cite{novotny2006,kuhn06}  The strong dependence of these effects
on fluorophore-metal separation has been used to increase both lateral and longitudinal resolution in confocal microscopy, with important
consequences for biological imaging.\cite{alschinger03}

Raman spectroscopy provides an enticing alternative to fluorescence, as it has many potential advantages and it is able to provide
high-resolution information with chemical specificity. As such, the technique is finding increasing application in biological
systems,\cite{hanlon00} as it allows rapid diagnosis using spectral deconvolution techniques such as principal component analysis. However, the
very weak intrinsic cross-section of the Raman scattering process (some 14 orders of magnitude less than fluorescence cross-sections) makes it
impractical for many biological applications that require low laser powers and short integration times. This Raman signal can be enhanced by
many orders of magnitude via surface enhanced Raman scattering (SERS) in the vicinity of noble metal nanoparticles.\cite{kneipp02} Devices that
utilise this effect can act as ultra-sensitive biosensors, with diverse applications in all areas of pathogen detection.\cite{zhang05}

Here, we report the results of high spatial resolution time-resolved measurements on individual self-assembled Ag nanoparticles. Using
fluorescence lifetime imaging microscopy (FLIM), we have observed localized enhancements in the emission intensity and additional fluorescence
lifetime components from vicinal fluorophores. By correlating the initial emission intensity and lifetime resulting from these modified decay
channels, we attribute these enhancements to a greater photon recycling rate due to coupling with surface plasmons. High resolution spatial maps
of Raman scattering show an enhancement in signal over the nanoparticles of more than 5 orders of magnitude.

\section{EXPERIMENTAL DETAILS}

\begin{figure}[tb] \begin{center} \includegraphics[width=16.5cm]{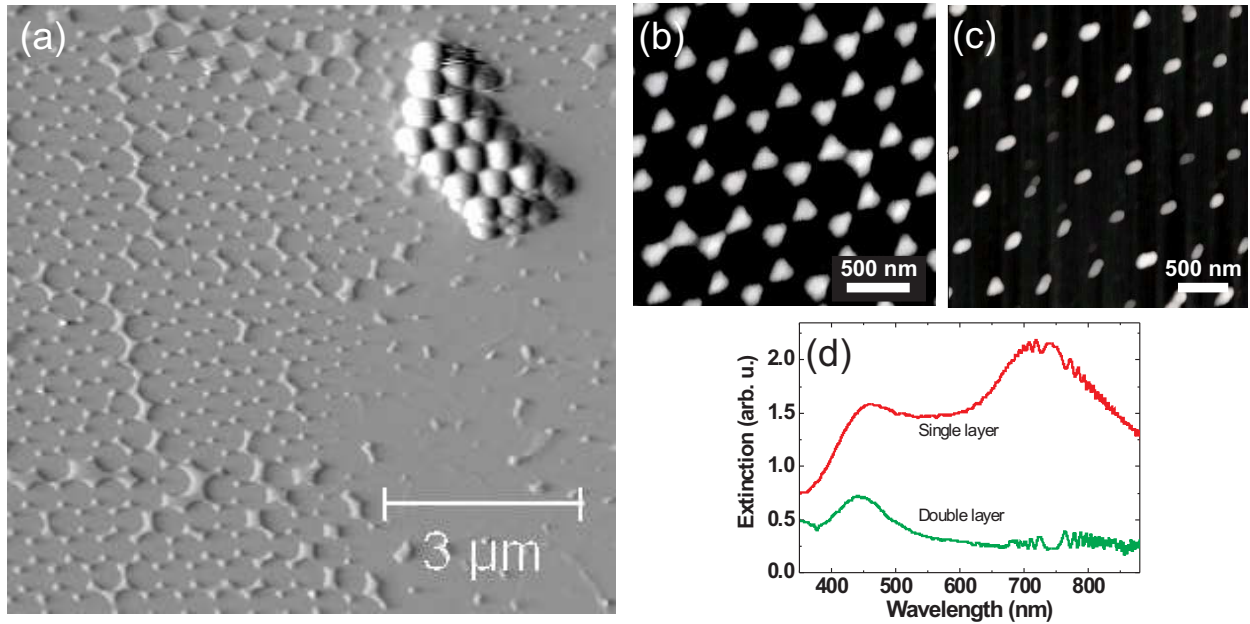} \end{center} \caption{ (a) AFM image of a nanoparticle array created
by a single layer of latex spheres. Some residual spheres still remain in the top right corner. (b) Zoom image from a region on the left side of (a), showing individual particles. (c) Nanoparticles formed from Ag deposition on a double layer of spheres. (d) Ensemble extinction spectra from single  and double layer Ag nanoparticle arrays, showing dipole and quadrupole plasmon resonances. The high-frequency noise is from interference in the glass slide.} \label{afm}
\end{figure}

Aqueous 500 nm diameter latex spheres were drop-cast onto a piranha treated glass slide, resulting in the formation of a monolayer / bilayer
close-packed lattice as the water evaporated.\cite{vanduyne2001} A 0.5 nm layer of chromium was then deposited onto the slide by thermal
evaporation in a vacuum chamber; this increases the adhesion of silver which was subsequently deposited to a thickness of 25 nm. The latex
spheres were removed by sonication in chloroform, leaving a continuous periodic array of Ag nanoparticles over a typical area of 20 mm$^2$.
Figure \ref{afm}(a) shows an atomic force microscopy (AFM) image of a boundary region of nanoparticles, formed from a single layer of latex
spheres (left hand side). There is a small region of residual latex spheres in the top right corner. Figure \ref{afm}(b) shows individual
particles within the array, which have a base length of $\sim$150 nm and height of 25 nm. In some regions, different structures form due to Ag
deposition on double layers of spheres,\cite{vanduyne2001} as shown in Fig.\ \ref{afm}(c).

Extinction spectroscopy (Perkin-Elmer Lambda 800) was used to determine the local surface plasmon resonances (LSPR) of the Ag nanoparticles.
Figure \ref{afm}(d) shows ensemble extinction spectra from single and double layer regions. The single layer structures have two strong
absorption resonances corresponding to dipole (720 nm) and quadrupole (460 nm) LSPR modes,\cite{schatz2003,nelayah2007} whereas the double layer
structures have only one peak at 440 nm due to their ellipsoidal geometry.\cite{vanduyne2001}

Rhodamine 6G (R6G) dye was deposited onto the nanoparticles by vacuum sublimation to produce a uniform coverage of sub-monolayer thickness. The
sample was prepared for high resolution optical measurements by applying a thin layer of index-matching polymer solution (Mowiol 4-88) to a
glass slide and affixing this on top of the nanoparticle array. During this procedure the R6G was incorporated into the polymer layer, creating
a homogeneous distribution of fluorophores across the sample with a thickness of several microns. This was verified by spectroscopic analysis of
excess polymer from the edge of the slide.

Fluorescence lifetime and intensity images were obtained using a scanning confocal epifluorescence microscope  (Leica TCSP2, 100x oil objective,
1.4 NA) with a synchronous time correlated single photon counting module (Becker and Hickl SPC-830). Excitation was with a 488 nm
continuous-wave (CW) Ar$^{+}$ laser (intensity) or 467 nm 20 MHz pulsed diode laser (lifetime), at a power far below fluorescence saturation in
both cases. Raman spectra and images were obtained using a Renishaw spectrometer with a high precision Prior H101 scanning stage. To reduce fluorescence background, excitation was at 752 nm using a tunable CW Ti:Sa laser with a 100x oil objective.

\subsection{Fluorescence Enhancement}
\label{sec:fluo}

\begin{figure}[tb]
\begin{center}
\includegraphics[width=16cm]{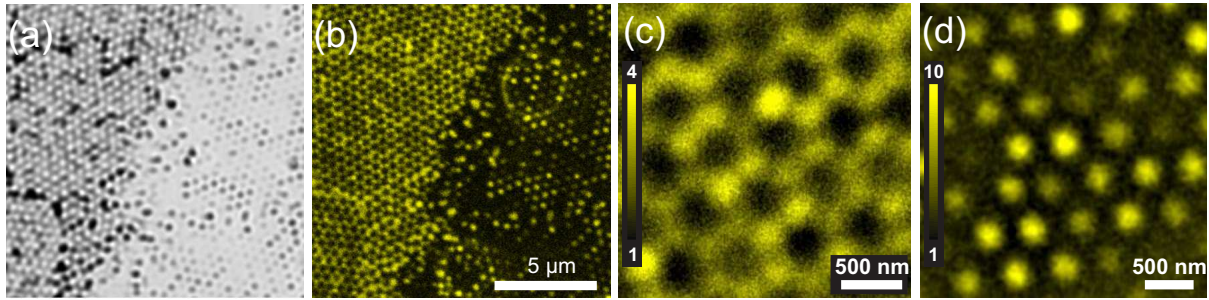}
\end{center}
\caption{ (a) Transmitted white light image of a boundary region between single layer (left) and double layer (right) nanoparticles. (b) Corresponding confocal  fluorescence intensity map of R6G, for the same region. (c) and (d) High resolution fluorescence maps from single and double layer regions similar to Fig.\ 1(b) and (c), respectively.}
\label{fluorescence}
\end{figure}

Figure \ref{fluorescence}(a) is the transmitted white light image of a boundary region between  nanoparticles formed by a single layer of latex spheres (left) and a double layer (right). The corresponding confocal R6G fluorescence signal detected from this region is shown in Fig.\ \ref{fluorescence}(b). High resolution scans from this area are shown in Figs.\ \ref{fluorescence}(a) and (b) for the single and double layer nanoparticles, respectively. These maps show highly localized fluorescence enhancement from molecules close to nanoparticles. Spectral analysis has verified that the emission originates from R6G and is not caused by photoluminescence from the silver nanoparticles or scattered laser light. The R6G fluorescence enhancement measured from the nanoparticles, relative to that from the glass, is approximately fourfold and tenfold for (c) and (d),
respectively. The actual enhancement in the vicinity of the nanoparticles will be much larger than that measured: enhancement is only expected
to occur over a very small range of metal-fluorophore distances ($\sim$20 nm),\cite{novotny2006} hence there is a significant unmodified
fluorescence background from the other fluorophores in the excited confocal volume. This is discussed further below.

There are two main causes of fluorescence \emph{enhancement} from molecules close to metal nanoparticles: when optically excited at its plasmon
resonance, a nanoparticle can dramatically enhance the local electromagnetic field intensity, resulting in an increase in a molecule's
excitation rate.\cite{schatz2003,nelayah2007} The excited-state molecular dipole can also couple with surface plasmon electrons in the metal
creating an additional radiative decay channel.\cite{barnes1998,lakowicz2005,gheddes2007}

\subsection{Lifetime Modification}
\label{sec:lifetime}

For CW measurements it is not possible to separate the contributions to the fluorescence enhancement arising from the modified excitation and
decay channels; however, this is not the case with time-resolved measurements. The quantum efficiency $Q$ gives the probability of an excited
state molecule decaying to a lower state by photon emission. In free-space $Q_{0}=\Gamma \tau_{0}$, where $\Gamma$ is the radiative emission
rate, $\tau_{0} = (\Gamma + k_{\mathrm{nr}})^{-1}$ is the total lifetime of the molecule, and $k_{\mathrm{nr}}$  is the sum of the non-radiative
decay rates. The presence of a metal creates an additional decay channel $\Gamma_{\mathrm{m}}$ for an excited molecule, so the total radiative
decay rate becomes $\Gamma^{\prime}=\Gamma+\Gamma_{\mathrm{m}}$. The modified fluorescence quantum yield $Q_{\mathrm{m}}$ and lifetime
$\tau_{\mathrm{m}}$ are then related by $Q_{\mathrm{m}}=\Gamma^{\prime}\tau_{\mathrm{m}}$, and $\tau_{\mathrm{m}} = (\Gamma^{\prime}+
k_{\mathrm{nr}}^{\prime})^{-1}$, where $k_{\mathrm{nr}}^{\prime}$ is the modified total non-radiative decay rate.

\begin{figure}[tb]
\begin{center}
\includegraphics[width=14cm]{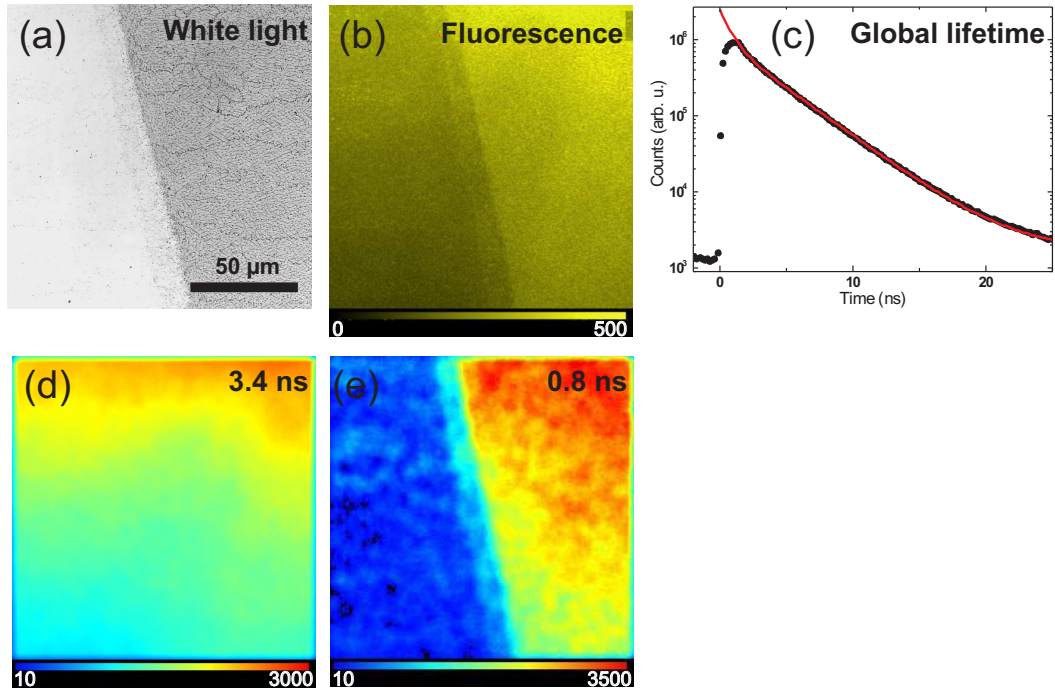}
\end{center}
\caption{Fluorescence lifetime analysis for a glass / nanoparticle boundary: (a) Transmitted white light. (b) Corresponding time-integrated fluorescence intensity of R6G. (c) Total fluorescence decay transient integrated over the whole region; the line is a biexponential fit to the data. (d) Spatial map of the preexponential intensity for the 3.4 ns lifetime component. The gradual variation in intensity is due to non-uniform illumination. (e) as (d) for the 0.8 ns lifetime component.} \label{lifetime1}
\end{figure}

The effects of Ag nanoparticles on the R6G fluorescence lifetime were investigated for regions of single-layer and double-layer nanoparticles. Figure \ref{lifetime1}(a) is the transmitted white light image of a glass \ nanoparticle boundary, with single-layer nanoparticles on the right. Fig.\ \ref{lifetime1}(b) shows the corresponding time-integrated fluorescence intensity from R6G. After excitation by a short laser pulse, the time evolution of the fluorescence intensity of a free-space single molecule is $I(t)= \alpha e^{-t /\tau_{0}}$. Figure \ref{lifetime1}(c) shows the spatially integrated raw decay transient from the region shown in \ref{lifetime1}(b); the data has been fitted with a biexponential decay.\footnote{An instrumental response of $\sim$60 ps has been included in all lifetime
analysis.} The spatial variations in the preexponential intensity of the two lifetime components are shown as FLIM maps in Figs.\ \ref{lifetime1}(d) and (e). The 3.4 ns component has a gradual variation in intensity due to non-uniform illumination, but is otherwise continuous across the nanoparticle boundary. However, the 0.8 ns short lifetime component only occurs over the nanoparticles.

\begin{figure}[tb]
\begin{center}
\includegraphics[width=17.1cm]{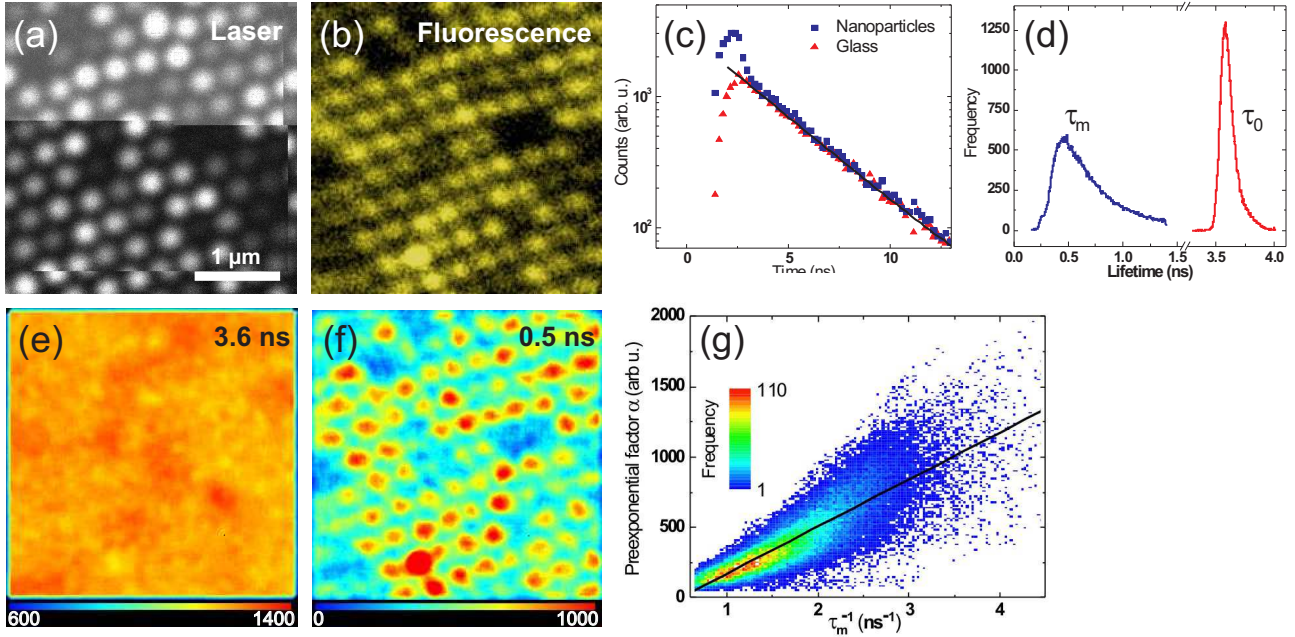}
\end{center}
\caption{Fluorescence lifetime analysis for a region of double-layer nanoparticles:  (a) Reflected laser intensity and (b) integrated fluorescence, from a small region similar to Fig.\ 2(d). (c) Raw decay transients from R6G on glass and nanoparticles; the straight line is a linear fit to the glass data. (d) Histogram of lifetime components ($\tau_{\mathrm{m}}$, $\tau_{0}$) obtained from biexponential pixel fits of the scan area. (e) Spatial map of the preexponential intensity for the 3.6 ns $\tau_{0}$ component, showing homogeneity across the region. (f) as (e) for the 0.5 ns $\tau_{\mathrm{m}}$ component. Note the difference in the intensity scales. (g) Pixel-by-pixel correlation between spatial maps of
$\tau_{\mathrm{m}}$ and associated $\alpha$ values, over the whole scan region. The color bar gives the binned frequency, and the line is a linear best fit.} \label{lifetime}
\end{figure}

To investigate this in more detail, similar measurements were made for a small area of double-layer nanoparticles: Figs.\ \ref{lifetime}(a) and (b) show the reflected laser and fluorescence intensity maps, respectively, for a scan area similar to that of Fig.\ \ref{fluorescence}(d). The spatially integrated decay transients from large areas of glass and nanoparticles are shown in Fig.\ \ref{lifetime}(c). In both cases there is an
identical long-lived component originating from unmodified fluorophores in the polymer layer; this has a monoexponential decay $>$3 ns and
implies that there are no significant concentration-induced non-radiative channels.\cite{leitner85} A FLIM map of this region was acquired for a 30 min integration period. Bins of $9\times9$ pixels ($20\times20$ nm) were used to give decay transients
with total preexponential counts of $\sim$$10^3$. Biexponential fits were then applied to give spatially resolved maps of constituent lifetimes
$\tau_{i}$ and their corresponding intensities $\alpha_{i}$. A histogram of the extracted lifetime components is shown in Fig.\ \ref{lifetime}(d): this comprises a sharply peaked component
$\tau_{0}$ from the unperturbed fluorophores, and a broader fast component $\tau_{\mathrm{m}}$ arising from fluorophores with modified decay
rates. Figures \ref{lifetime}(e) and (f) show maps of the preexponential intensities for the two modal lifetime components; the 0.5
ns component is localized (resolution limited) around the nanoparticles, whereas the 3.6 ns component has a uniform intensity (standard
deviation $< 2 \%$) consistent with the homogeneous coverage of fluorophores (c.f.\ Figs.\ \ref{lifetime1}(d) and (e)).

For this sample, the R6G (in polymer) layer thickness is comparable to the axial confocal excitation depth; hence, fluorophores at different
distances from the nanoparticles will contribute differently to the total signal measured. Molecules in very close proximity to a metal show
strong fluorescence quenching due to dominant non-radiative energy transfer.\cite{dulkeith02} We have taken measurements on samples without the
additional polymer layer; these show almost complete quenching of the R6G fluorescence over the Ag nanoparticles, and an instrumental-response
limited lifetime. This strong distance-dependent weighting means that, for the current sample, the \emph{enhanced} fluorescence with a lifetime
$\tau_{\mathrm{m}}$ is from a thin shell of molecules around the nanoparticles where the competing quenching and enhancement mechanisms result
in a maximum net emission intensity.\cite{huang,novotny2006}

An aqueous R6G molecule has $Q=0.95$; hence, we assume that fluorophores showing a maximally enhanced emission intensity have $Q_{\mathrm{m}}
\approx 1$, so $\Gamma^{\prime} \approx \tau_{\mathrm{m}}^{-1}$. The preexponential factor $\alpha$ depends on several fixed experimental
parameters and also directly depends on the radiative decay rate $\Gamma^{\prime}$.\cite{gerber07} Thus, we expect a direct correlation between
$\alpha$ and the measured decay rate. Figure \ref{lifetime}(g) shows a pixel-by-pixel correlation between the spatial maps of the
$\tau_{\mathrm{m}}$ lifetime and their associated $\alpha$ values, extracted from biexponential fits over the region of nanoparticles as described
above. The data have been binned and the resulting frequency at each point is shown by the color bar. Scatter plots of $\chi^2$ versus $\alpha$
and $\tau$ were created from all biexponential fits; these show no fitting related bias. The data in Fig.\ \ref{lifetime}(g) are well fit by a
linear regression, in agreement with the above discussion. A large number of fluorophores show an order of magnitude increase in decay rate
relative to the unmodified background ($\tau_{0}^{-1}=0.28$ ns$^{-1}$), with a commensurate enhancement in initial intensity, as shown in Fig.\
\ref{fluorescence}(d). These results are consistent with reports of single-molecule lifetime modification using a scanning nanoantenna.\cite{kuhn06}
In CW measurements, the nanoparticle-induced enhancement in integrated intensity cannot be accounted for by an increase in $Q$. Instead it
originates primarily from a greatly increased photon recycling rate for fluorophores with a strong radiative coupling to surface plasmons.

\subsection{SERS}
\label{sec:sers}

\begin{figure}[]
\begin{center}
\includegraphics[width=15cm]{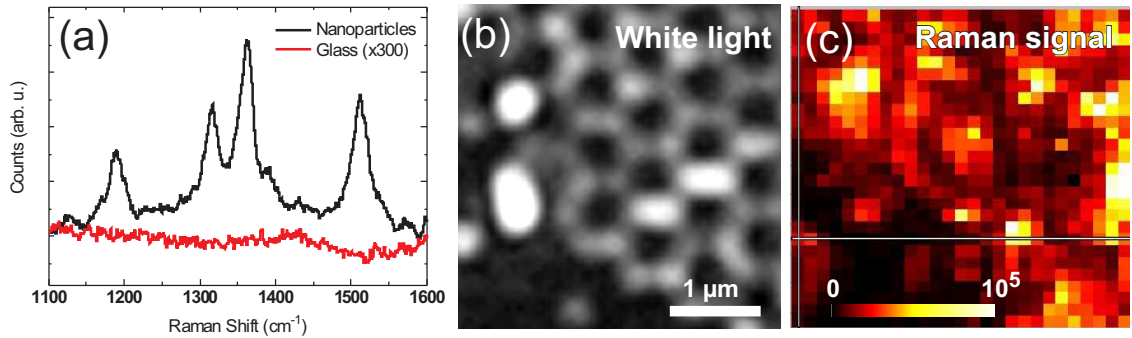}
\end{center}
\caption{ (a) Raman spectra from R6G on nanoparticles and glass. (b) Reflected white light image of nanoparticle array. (c) Intensity map of the
1500 cm$^{-1}$ Raman line, from a region similar to (b).} \label{raman}
\end{figure}

Figure \ref{raman}(a) shows the Raman spectra from R6G on a region of single-layer nanoparticles (2 s acquisition time) and from glass (10 min acquisition time). As no signal is obtained from the R6G on glass, it is not possible to calculate an absolute value for the enhancement factor; however, a lower limit of $10^5$ can be estimated using the noise level in the glass spectrum. This has been verified by obtaining high resolution spatial maps of the Raman signal over the nanoparticles. Figure \ref{raman}(c) shows the integrated intensity under the 1500 cm$^{-1}$ Raman line, for a region of nanoparticles similar to \ref{raman}(b). 
The huge localized enhancement in Raman signal is due to the increase in the local electric field intensity around an individual nanoparticle:\cite{kneipp02} the total SERS enhancement has an $E^4$ dependence on electric field due to enhancement of the incident laser field and enhancement of the emission field at the Raman frequency. This means that localized differences in the surface plasmon modes result in the creation of optical ``hot spots", as has been previously reported for metal films.\cite{gresillon99}

\section{CONCLUSION}

In conclusion, we have demonstrated highly localized modification of fluorescence intensity and lifetime on ordered arrays of Ag nanoparticles.
FLIM maps of R6G fluorescence show an additional plasmon-induced radiative decay channel for a thin shell of molecules around individual
nanoparticles. This leads to an order of magnitude increase in ensemble photon recycling rate with a proportionate enhancement in fluorescence
intensity. The strong distance dependence of these effects suggests that metal enhanced fluorescence may offer a means of selectively mapping
the location of specific target molecules within a larger excitation volume, such as fluorophore-tagged proteins in cell membranes. Furthermore, spatially resolved maps of Raman scattering from R6G show localized enhancement from individual nanoparticles of more than five orders of magnitude. This has important implications for the development of ultra-sensitive biosensors that require a high degree of chemical specificity.

\section{ACKNOWLEDGEMENTS}

The authors would like to thank F. Festy and K. Suhling for helpful discussions. This work was supported by the EPSRC (UK).

\end{document}